\documentclass[twocolumn,showpacs,floatfix,amsmath,amssymb,prl]{revtex4}
\usepackage{graphicx}
\usepackage{dcolumn}
\usepackage{bm}
\def\be{\begin{equation}} 
\def\ee{\end{equation}}
\def\bea{\begin{eqnarray}} 
\def\eea{\end{eqnarray}}
\def \line{\hbox to \hsize}    
\def\frac #1#2{{#1\over #2}}

\def\psid{\psi^{\dagger}}

\def \ket #1{{\vert #1\rangle}}

\def\1{\mbox{\bf 1}}
\newcommand{\comment}[1]{}

\begin{document}

\title{Topological superfluids with time reversal symmetry}

\author{Rahul Roy}

\affiliation{ Department of Physics and Astronomy, McMaster University\\Hamilton, Ontario, Canada L8S 4M1}   

\begin{abstract}
  It is shown that superfluids in two and three dimensions which have
  time reversal invariant ground states have phases which are
  distinguished by a topological invariant. Further, it is shown that
  the B-phase of $^3$ He is a superfluid in the non-trivial
  topological class.  Superfluids in the non-trivial topological class
  are shown to have gapless edge states and support various kinds of
  vortices with zero energy modes localized in their cores. Some of
  these vortices have non-abelian statistics.
\end{abstract}

\maketitle

There has been a considerable amount of interest in phases
characterized by a topological invariant (topological phases). The
most well known examples of such phases are the quantum Hall
states~\cite{thouless1982qhc,niu1985qhc,kohmoto1984tiq}.  Chiral
superconductors and superfluids are another class of systems which are
characterized by a topological invariant~\cite{volovik97}. A droplet
of $^3$He in the A-phase~\cite{leggett1975tdn} is an example of a 2D
chiral superfluid while strontium ruthenate is thought to be a chiral
superconductor~\cite{mackenzie2003ssp}. Chiral superconductors and
superfluids have a number of interesting properties. A neutral chiral
superfluid has a macroscopic current which runs along the edge of a
bounded sample. In a superconductor, these currents are screened but
still substantial~\cite{furusaki2001she,stone2004eme}.  Chiral
superconductors and superfluids have gapless edge states. They also
support exotic defects called half quantum vortices which carry half a
quantum of flux. Each half quantum vortex carries a Majorana fermion
and a pair of Majorana fermions is equivalent to a Dirac fermion.  A
system of $2n$ vortices thus has a $2^{n}$ dimensional vector space of
degenerate ground states.  The zero energy Majorana fermions are
stable since they cannot couple to local operators.  When the vortices
are braided around each other, the vector corresponding to the ground
state is rotated in the $2^{n}$ dimensional vector space and this
phenomenon is known as non-abelian
statistics~\cite{read2000psf,ivanov2001nas,stone2006fra}.

Chiral superconductors and the quantum Hall states break time reversal
symmetry.  Until recently, examples of known topological phases were
restricted mostly to systems whose ground states broke time reversal
symmetry. In this paper we study superfluids with unbroken
translational symmetry whose ground states are time reversal invariant
in two and three dimensions. We find that the ground states of these
systems in two and three dimensions can be classified by means of a
topological invariant and that superfluids in the non-trivial
topological class have a number of interesting properties. These systems
have robust gapless edge states which are stable against perturbations
that do not break time reversal symmetry. They can also support exotic
defects which have zero-energy Majorana fermions and can lead to
non-abelian statistics.

There has been considerable recent work on materials and systems with
time reversal symmetry which have a non-trivial topological $Z_2$
invariant and some of these are believed to have been experimentally
detected~\cite{konig2007qsh,Hasan08}. These systems are generically
called quantum spin Hall systems~\cite{koenig-2008}. We show that the
$B$ phase of superfluid Helium-3 can be identified as a ``topological
superfluid'' in the non-trivial topological class.

For the rest of this work, unless explicitly stated, by superfluids,
we mean superfluids with unbroken translation symmetry. Lattice
superconductors which are not invariant under the full translational
group have previously been studied in an earlier work~\cite{roy-2006}.
In Sec.~I, we study and obtain a topological classification of
superfluids in two and three dimensions. In Sec.~II, we study the
edge states in these superfluids. In Sec.~III, we present examples
of superfluids in the non-trivial topological class. In Sec.~IV, we
consider exotic vortices and their statistics in these systems. 

\section{I. Topological superfluids with time reversal symmetry}
\label{sec:topol-superfl}

 The mean field BdG 
Hamiltonian which characterizes a superfluid may be written in the form: 

\begin{eqnarray}\label{eq:3}
   H_{BdG} = \int \,d^d k\, \left[ \psid(\bm{k})\, \hat{H}
     \,\psi(\bm{k}) \right],
\end{eqnarray}
where 
  \begin{eqnarray}\label{eq:4}
  \hat{ H} & = & \left(
   \begin{matrix} \hat{h}(\bm{k}) & \hat{\Delta}(\bm{k}) \\  \hat{\Delta}^{\dagger} (\bm{k}) & 
   -\hat{h}^{T}(-\bm{k}) \end{matrix} \right).  
   \end{eqnarray} 

   Here $\hat{h}$ represents the single particle Hamiltonian, $\Delta$
   the order parameter that characterizes superconductivity and $\psi,
   \psid$ are the fermionic operators in the two component Nambu
   formalism.  The BCS ground state wavefunction of the superfluid is
   annihilated by operators of the form $ \gamma(k) = \sum_{\alpha=
     \uparrow,\downarrow}u_{\bm{k},\alpha}\psi_{\bm{k},\alpha} +
   v_{\bm{k},\alpha}\psid_{-\bm{k},\alpha}$ where $
   {(u_{\bm{k},\alpha}, v_{\bm{k},\alpha} )}^T $ is a negative energy
   eigenvector of the matrix $H$.

 As long as the system has a
   bulk gap, at each point in momentum space, there are two
   eigenvectors of this matrix, say $e_1(\bm{k})$ and $ e_2(\bm{k})$
   with negative eigenvalues. As $k \rightarrow \infty$, the two
   dimensional vector space spanned by these states, which we denote
   by $V(k)$ goes to a fixed two dimensional space~\footnote{This is
     based on the assumptions that $h(\bm{k})$ becomes a function
     only of the magnitude of $k$ and that $\Delta(k)$ vanishes as $k
 \rightarrow \infty $}. As far as the topology of the ground state
   wavefunction is concerned, the base space which is momentum space
   is thus a sphere, $S^n $ obtained by the one point compactification
   of $R^n$, where n is equal to 2 or 3 in the cases of our interest.
   
   When the ground state wavefunctions has time reversal symmetry, the
   eigenvectors of $\hat{H}$ at $\bm{k}, -\bm{k}$ are not independent.
Time reversal symmetry requires that if $u$ is an eigenstate at $\bm{k}$,
then $\Theta u$ is also an eigenstate with the same energy at $-\bm{k}$.
While the two dimensional vector space consisting of the negative
energy eigenvectors can be written in terms of a continuous basis
locally in momentum space, finding a global continuous basis is not
always possible. The ground state thus defines a twisted two
dimensional vector bundle on $S^2 $ or $S^3$ as the case may be.
   
\subsection{Superfluids in 2D}   
Let us first study the topology of the ground state wavefunction of a
two dimensional superfluid following~\cite{roy2006zcq}. We divide
$S^2$ into two patches, E and E' where $ E = \{ \bm{k} \ni k < 2\}$
and $E' = \{\bm{k}\ni k >0.5 \}$.  On each of these patches we define
a basis of two continuous orthogonal vector functions,
$\ket{v_{1,i}(\bm{k})}, \ket{v_{2,i}(\bm{k})} \in V(\bm{k}) $ where
the index i is a label for the patch E or E'. We represent the vector
$ v_i(\bm{k}) = c_1 v_{1,i}(\bm{k}) + c_2 v_{2,i} (\bm{k}) $ as the
spinor ${ (c_1, c_2 )}^T$.  The operation of time reversal in this basis
then corresponds to the operator $i\sigma_2 K_0 \Phi$ where $K_0$ is
the complex conjugation operator and $\Phi$ is the involution operator
that takes vectors at $\bm{ k}$ to vectors at $-\bm{k}$.  If a vector
$v$ is represented by the spinors $v_E$ and $v_{E'}$ in the two
different bases defined above, then $v_E$ and $v_{E'}$ can be related by
means of a transformation matrix U such that $ v_E = U v_{E'}$ where $U$ 
is an element of the unitary group, $ U(2)$.  The
transition function, $U(\bm{k})$, which glues the vector
representations $v_E(\bm{k}), v_{E'} (\bm{k})$ on the two patches, on
the circle $ T =\{\bm{k} \ni k =1 \} \in E \bigcap E'$ can be written
as $ e^{i\alpha(\bm{k})I}e^{i\bm{n}(\phi).\bm{\sigma} \omega(\bm{k})}$.
 Then time reversal invariance leads to the condition:
\begin{eqnarray}
 \label{tricondition} U(-\bm{k}) = e^{-i\alpha(\bm{k})I}e^{i\bm{n}(\phi).\bm{\sigma} \omega(\bm{k})}. 
\end{eqnarray}

 Then, as shown in previous work~\cite{roy-2006}, $U_{2D}$, the set of all transition
 functions for these ground states in two dimensions consists of two
 topologically distinct classes of transition functions. The trivial
 class corresponds to the case when it is possible to find a globally
 continuous basis for the vector space spanned by $e_1(k),
 e_2(k)$. The non trivial $Z_2$ phase corresponds to the case when the
 transition function cannot continuously be deformed to unity. A
 simple formula for the $Z_2$ invariant can be written down when the
 wavefunctions $e_1 (k), e_2 (k)$ have well defined Chern
 numbers. When this is not the case, one can still always find a set
 of two orthogonal ``wavefunctions'' $ e_1'(\bm{k}),e_2 '(\bm{k}) \in
 V(k)$ for which well defined Chern numbers exist. The $Z_2$ invariant
 can then be expressed in a particularly simple form in terms of these
 Chern numbers. It is given by $ |c_n|\mod 2 $ where $c_n$ is
 the Chern number of either of these functions.

\subsection{Superfluids in 3D}
Now consider a superfluid in three dimensions. We again divide
momentum space, which is now topologically equivalent to the three
sphere $S^3$ for our purposes, into two patches $ E = \{ \bm{k} \ni k
< 2\}$ and $E' = \{\bm{k}\ni k >0.5 \}$. As before, on each of these
patches we define a basis of two continuous orthogonal vector
functions, $\ket{v_{1,i}(\bm{k})}, \ket{v_{2,i}(\bm{k})} \in V(\bm{k})
$ where the index i is a label for the patch E or E'. 

The transition function, $U(\bm{k})$ which glues the vectors on the
two patches, $v_E(\bm{k}), v_{E'} (\bm{k})$ on the two sphere $ T = \{
\bm{k} \ni k =1 \} \in E \bigcap E' $ is again an element of $U(2)$
and can be written as $ e^{i\alpha(\bm{k})I}e^{i
  \bm{n}(\phi).\bm{\sigma} \omega(\bm{k})} $. Time reversal symmetry
results again in the constraint given by Eq.~(\ref{tricondition}).
Let C be any great circle on T. Then the function obtained by
restricting the domain of $U(\bm{k})$ to T is a function which belongs
to the class of two dimensional transition functions $U_{2D}$. This
thus defines a projection from the class of transition functions on
$S^2$ for the three dimensional case, which we call $U_{3D}$, to the
class of transition functions for the two dimensional case, $U_{2D}$.
As previously discussed, there are two topologically inequivalent
classes in $U_{2D}$.
       
Further, it is well known that $\pi_2 ( U(2)) = \pi_2 (SU(2))=0
$.
 This implies that the class of transition functions in 3D is determined
by the topological class of the 2D transition function that it projects to.

Thus the topological invariant for the 3D superfluid may also be
evaluated using an extension of the simple Chern number formula from
the 2D case.  Let P be any two dimensional plane which maps onto
itself under time reversal and contains the origin and let $
e_1'(\bm{k}),e_2 '(\bm{k}) $ be a set of orthogonal wavefunctions
which span $V(\bm{k})$ at each point $\bm{k}$ in the plane, which map
onto each other under time reversal symmetry and for which well
defined Chern numbers exist.  Then, the 3D $Z_2$ invariant is simply $
|c_n|\mod 2 $, where $c_n$ is the Chern number of one of the
wavefunctions.

We have shown above that there are thus two distinct topological
phases of superfluids at the mean field level. This implies that if
the full Hamiltonian of the system is adiabatically perturbed, as long
as the ground state does not spontaneously break time reversal
symmetry, the ground state remains in one of two distinct phases,
unless there is a phase transition.

We note that superfluids in 2D have a single $Z_2$ invariant as do
lattice superconductors. However in 3D, lattice superconductors have
four $Z_2$ invariants~\cite{roy-2006}, while superfluids with unbroken
translational symmetry have a single $Z_2$ invariant. This invariant
corresponds to the fourth $Z_2$ invariant which is intrinsically three
dimensional in nature and corresponds to the invariant in insulators
and lattice superconductors which determines whether the system is in
the strong or weak topological class.

\section{II. Edge states in topological  superfluids}
\label{sec:ii.-edge-states}
\subsection{Superfluids in 2D}
A chiral 2D superconductor and a quantum Hall insulator are both
characterized by the same topological number, namely the first Chern
number~\cite{volovik97,kohmoto1984tiq}.  The connection between the
Chern number and the presence of robust edge states is well known in
the context of the integer quantum Hall
effect~\cite{halperin1982qhc,hatsugai1993cne}.  Further, it is also
known that these edge states exist in chiral superconductors and
superfluids. This follows both from the Chern number -edge state
connection and also from explicit calculations using the BdG
Hamiltonian~\cite{stone2004eme}.  The gapless edge excitations in the
case of a chiral superfluid/superconductor are Majorana fermions or
equal linear combinations of particles and holes.  In the case of
insulators with time reversal symmetry, the $Z_2$ invariant determines
whether robust edge states exist or not. A superconductor whose ground
state respects time reversal symmetry can not have net charge currents
along the edge. Nevertheless since the ground state of a non-trivial 2D
$Z_2$ superconductor may be adiabatically continued to a product state
of two wavefunctions each of which has the same form in position space
as the ground state wavefunction of a chiral superconductor with an
odd Chern number, it follows that the TR invariant superconductor of
the non trivial $Z_2$ class has a robust pair of edge states.  When
the Chern number is one, there is precisely one set of edge states and
a single pair of Majorana fermions at zero energy at the edge. This pair of
Majorana edge states is stable as long as time reversal symmetry is preserved
because no operator that is invariant under time reversal symmetry can have
a non-zero matrix element between the two states~\cite{xu:045322,wu2006hle}. 
  
\subsection{Superfluids in 3D}
We now study the 3D case.  Consider a semi infinite three dimensional
superfluid with periodic boundary conditions in any two directions,
say in the x and y directions and which exists in the region $ z< 0 $.
The momentum variables $k_x, k_y $ are then good quantum numbers.  The
Hamiltonian $H(\alpha k_x + \beta k_y=0,z)$ represents a 2D superfluid
Hamiltonian which from our previous discussion is a 2D TR invariant
Hamiltonian in the non-trivial $Z_2$ class.  It therefore follows that
the eigenstates of this Hamiltonian has a set of robust edge states
with a 1D Dirac spectrum. Since this is true for arbitrary
$\alpha,\beta$, it follows that the eigenstates of the Hamiltonian in
general have a 2D Dirac energy spectrum at low energies. While the
spectrum is Dirac-like, the eigenstates are Majorana rather than Dirac
fermions. There is a single pair of zero energy Majorana fermions at the
edge corresponding to the $k_x = k_y = 0$ state.
 More generally when there are open boundary conditions, there
are two surfaces. From the above argument, it is expected that each one will
have a single pair of zero energy Majorana fermions~\footnote{The
  energy splitting induced by the finite distance between the edges
  falls of exponentially with the distance}.

\section{III. Examples of topological superfluids}
In this section, we shall present some examples of states in two and
three dimensions that are in the non-trivial $Z_2$ class. Since these
states are triplet states, it is useful to introduce the $\bm{d}$-vector
notation. The order parameter of a general triplet
superconductor/superfluid can be written in the
form~\cite{leggett1975tdn}:
\begin{eqnarray}
 \Delta =\Delta_0 \left( \begin{matrix}  i d_y -d_x & d_z \\ d_z & i d_y + d_x \end{matrix} \right)
\end{eqnarray}
where in the absence of textures, defects and edges, $\bm{d}$ is a
function only of $\bm{k}$.
  
The state whose d-vector is given by 
\[
\bm{d(k)}=i(k_x+ik_y)\bm{\hat{y}}
\]
 is an
example of a chiral $p+ip$ superconductor/superfluid.  On the other hand, the
state with 
\begin{eqnarray}\label{eq:5}
\bm{d(k)}=k_x\bm{\hat{x}}+k_y\bm{\hat{y}}
\end{eqnarray}
 is an example of a
superfluid with a non-trivial $Z_2$ invariant~\footnote{This is a two
  dimensional version of the Balian-Werthamer state.}. Since the spins
decouple for such a state, the up and the down spins may be considered
separately.  The wavefunction for the up/down-spins are identical to
those of a chiral $p_x - ip_y / p_x+ip_y$ superfluid. Thus, the form
of the wavefunctions and their spectrum for chiral superfluids in
rectangular and circular geometries with an edge~\cite{stone2004eme}
may be used to deduce the wavefunction for the ground state of the
above system, but shall not be presented explicitly here.  It follows
from the analogy with chiral superfluids, that there is a single pair
of stable zero energy Majorana edge states as previously stated in Sec.~II.

The Balian Werthamer state which is believed to exist in the B-phase of
$^3$He has the description :
\begin{eqnarray}
\bm{d(k)} = \bm{k}
\end{eqnarray}
This is a gapped 3D superfluid which is time reversal invariant. To
determine its topological invariant, we consider the plane $k_z =0 $
in momentum space.  The order parameter then corresponds to the 2D
superfluid given by Eq.~(\ref{eq:5}) which is in the non-trivial $Z_2$
class of 2D supefluids. From the results of Sec.~I, it follows
therefore that $^3$He - B is a superfluid in the non-trivial $Z_2$
class. 

Indeed, as anticipated in Sec.~II, the
superfluid state when confined to a boundary has gapless edge
states. Since the BdG Hamiltonian for this state is invariant under
simultaneous rotations in spin and orbital space, one may consider
without loss of generality a three dimensional sample confined with an
infinite wall at y=0. We further consider the set of states with $k_z
=0$. The resulting system can be regarded as the sum of two two
dimensional chiral superfluids with opposite spin and
chirality~\footnote{More precisely, the Hamiltonian for these states
  is a direct sum of up and down spin components which are analogous
  to the Hamiltonians of chiral spinless superfluids}. The edge states
of a chiral 2D superfluid in a rectangular geometry were analyzed in
Ref.~\cite{stone2004eme}. It was found that the spectrum was linear in
the momentum component parallel to the edge and that there was
precisely one zero energy Majorana mode. It follows that the confined
3D superfluid considered here has a spectrum $E\propto \sqrt{k_x ^2 +
  k_z ^2}$ which is 2D Dirac-like.

  \section{IV. Exotic vortices and non-abelian statistics}

  The 2D topological superfluid state whose order parameter is given
  by Eq.~(\ref{eq:5}) supports many kinds of exotic vortices. In the
  vortices that we shall be interested in, the order parameter takes
  the form:
\begin{eqnarray}\label{eq:1}
 \Delta (r,\theta) =\Delta_0(r) \left( \begin{matrix}e^{i\phi_{-}(\theta)} ( i d_y -d_x) & 0
 \\ 0 &e^{i\phi_{+}(\theta)} (i d_y + d_x )\end{matrix} \right).
\end{eqnarray}
Here $r, \theta$ are the polar coordinates in the vortex and
$\phi_{\pm}$ are the phase windings of the different components of the
order parameter. When $\phi_{+}=\theta, \phi_{-}=0$ or vice versa, the
defect is a half quantum vortex. In a superconductor such a defect
would carry half a quantum of flux. Only one of the spin components
are involved in the low energy physics. We may therefore consider a
spinless chiral superconductor with a full quantum vortex, which as
pointed out in Ref.~\cite{stone2006fra} has a single zero energy
Majorana core state at the center of a vortex. Further the statistics
of these states is non-abelian. 
 
 When the system has both spin and orbital rotational symmetry
 the state given by Eq. (\ref{eq:5}) is degenerate with the set of states 
 \begin{eqnarray}\label{eq:2}
 \bm{d(k)}= R[\bm{k}] 
\end{eqnarray}
where $R[\bm{k}]$ represents the vector obtained by a rotation, $R$, of
the k-vector in the two dimensional momentum space plane. One can
therefore consider defects which are represented by
\begin{eqnarray}
  \label{eq:8}
  \phi_{+} =\pm \theta\,\,,\, \,\phi_{-}=\mp \theta 
\end{eqnarray}
These are spin-vortices which may be seen as arising from the slow
rotation of the $\bm{d(k)}$ in position space. The overall phase of
the order parameter does not vary and thus these defects preserve time
reversal symmetry. The spins are still decoupled and each spin
component can be regarded as a chiral superfluid which now has a full
quantum vortex but of opposite relative chirality.  Further, it
follows that there is a single pair of zero energy Majorana fermions
which are stable to local perturbations which do not break time
reversal symmetry. The statistics of these fermions for each
individual spin component is non-abelian~\footnote {When the
  statistics of the vortices as a whole are considered however, the
  statistics is abelian.} and this could possibly be harnessed for
quantum computation if the spins are kept decoupled.

The 3D superfluid previously discussed, which is described by the
equation: $\bm{d(k)}= \bm{k}$ is also degenerate with the set of states
$\bm{d(k)}=R[\bm{k}]$ where $R$ is any rotation in real space when the
system has both spin and orbital rotational symmetry.  Since $R$ is an
element of $SO(3)$ and $\pi_1 (SO)(3)= Z_2 $, the superfluid then
supports line defects which correspond to non-contractible paths in
$SO(3)$.  An example of such a vortex is 
\begin{eqnarray}\label{eq:6}
\bm{d(k,\theta)}=R(\bm{n},\theta)[\bm{k}]
\end{eqnarray}
where $\theta$ is the polar coordinate in the vortex and
$R(\bm{n},\theta)[\bm{k}]$ is the vector obtained by rotating $\bm{k}$
through the angle $\theta$ about the axis $\bm{n}$.  To determine the
low energy states, we apply periodic boundary conditions in the $z$
direction and set $\bm{n}=\bm{z}$. Then for the set of states $k_z
=0$, we see that the order parameter at the vortex has the same form
as the defect in the 2D superfluid considered in Eqs.~(\ref{eq:1}) and
(\ref{eq:8}). There is thus a stable pair of zero energy Majorana
states at the core of these vortices. The 3D superfluid state also
hosts analogs of the half quantum vortex. The discussion from the 2D
case can be easily carried over to the defects in the 3D superfluid
states.

The above discussion pertained to superfluids with unbroken
translational symmetry.  Lattice superconductors in two dimensions are
described by a single $Z_2$ invariant, while in 3D, they are described
by four such invariants. The fourth $Z_2$ invariant for
superconductors is analogous to the single $Z_2$ invariant for
superfluids discussed above. The discussions pertaining to the edge
states and the exotic defects presented above is also valid for
superconductors with a non-trivial $Z_2$ invariant in 2D and a non
trivial fourth $Z_2$ invariant in three dimensions.
\\

In conclusion, we have shown that superfluids which do not break
translational symmetry have two phases determined by topology. We also
showed that the non-trivial topological class of superfluids have a
number of interesting properties such as edge states and exotic
vortices which obey non-abelian statistics. We also provided examples
of such superfluids and showed that the B-phase of $^3$ He is an
example of a non-trivial topological superfluid. It is hoped that the
present work will spur further interest and work in these systems.

\subsection{Acknowledgments}
The majority of the work presented above was done during a visit to
the Indian Institute of Science (IISc.), Bangalore in Oct. - Dec. 2006
and I thank T. Senthil and IISc., Bangalore for their hospitality
during the visit.  This work was also partially supported by the Natural
Sciences and Engineering Research Council of Canada (NSERC), by the
Canadian Institute for Advanced Research and by the National Science
Foundation under Grant No.  DMR 06-03528.

I owe special thanks to Sheldon Katz and Michael Stone for very
educational and useful discussions on topology, to Shou-Cheng Zhang
for introducing me to the field of the quantum spin Hall effect and to
C. Wu, John Stack  and the Department of
Physics, UIUC for their help in various ways in making this and
earlier works on related topics possible. I also wish to thank
Catherine Kallin, Kumar Raman, Eduardo Fradkin, T. Senthil, Sung-Sik
Lee, Shou-Cheng Zhang, Charles Kane, Joel Moore and K. Shtengel for
useful discussions. I would also like to thank S. Chakraborty, A
Jaefari, A. Pushp, S. Sur and D. Ferguson for their assistance in preparing
this manuscript.



\end{document}